\documentclass[sigconf]{acmart}
\usepackage{ulem}
\AtBeginDocument{%
  \providecommand\BibTeX{{%
    \normalfont B\kern-0.5em{\scshape i\kern-0.25em b}\kern-0.8em\TeX}}}

\setcopyright{acmcopyright}
\copyrightyear{2024}
\acmYear{2024}
\acmDOI{XXXXXXX.XXXXXXX}

\acmConference[WSESE 2024]{International Workshop on Methodological Issues with Empirical Studies in Software Engineering}{2024}{Lisbon, Portugal}
\acmPrice{15.00}
\acmISBN{978-1-4503-XXXX-X/18/06}

\usepackage{graphicx}
\usepackage{todonotes}
\usepackage{soul}

\begin{document}

\title{The Role of Data Filtering in \\Open Source Software Ranking and Selection}

% \affiliation{%

\author{Addi Malviya-Thakur}
\affiliation{%
  \institution{University of Tennessee}
  \institution{Oak Ridge National Laboratory}
  \country{USA}
  }
\email{amalviya@vols.utk.edu}

\author{Audris Mockus}
\affiliation{%
  \institution{University of Tennessee, Knoxville}
  \institution{Vilnius University}
  %\city{Knoxville}
  \country{USA and Lithuania}}
\email{audris@utk.edu}


\begin{abstract}
Faced with over 100M open source projects, a more manageable small subset 
is needed for most empirical investigations. Over half of the research papers in leading venues investigated
filtering projects by some measure of popularity with explicit or
implicit arguments that unpopular projects are not of interest,
may not even represent ``real'' software projects, or that less
popular projects are not worthy of study.  However, such filtering
may have enormous effects on the results of the studies if and
precisely because the sought-out response or prediction is in any
way related to the filtering criteria. 

This paper exemplifies the impact of this common practice on research outcomes, specifically how  filtering of software projects on GitHub based on inherent characteristics affects the assessment of their popularity. Using a dataset of over 100,000 repositories, we used multiple regression to model the number of stars \textcolor{black}{--a commonly used proxy for popularity--} based on factors such as the number of commits, the duration of the project, the number of authors and the number of core developers. Our control model included the entire dataset, while a second filtered model considered only projects with ten or more authors. The results indicated that while certain characteristics of the repository consistently predict popularity, the filtering process significantly alters the relationships between these characteristics and the response. We found that the number of commits exhibited a positive correlation with popularity in the control sample but showed a negative correlation in the filtered sample. These findings highlight the potential biases introduced by data filtering and emphasize the need for careful sample selection in empirical research of mining software repositories. 
%~\cite{Mockus2008}. 
%The relationships obtained may have the opposite sign to those obtained with unfiltered data. %We use a simple example of modeling popularity of a project and demonstrate how filtering may lead to opposite results. It is extremely difficult to ensure that filtering does 
%Many published empirical studies are potentially flawed, and different results would have been obtained with different types of filtering, and that,
We recommend that empirical work should either analyze complete datasets such as World of Code, or employ stratified random sampling from a complete dataset to ensure that filtering is not biasing the results.  
\end{abstract}




\maketitle

%\todo[inline]{\url{https://docs.google.com/document/d/1mQ4mWR-kc77MuOWVwmg6iHJX5wKtZBcp91wZmubdPTs/edit?usp=sharing}}
\vspace{2mm}
\section{Introduction}
% \vspace{1mm}
Volumes of public data that could be extracted from open source projects that use version control and issue tracking systems have been widely exploited by research that seeks to better understand software development projects. Such studies increasingly use a large collection of projects to achieve some generality of the results. As it is extremely difficult to collect these data from all OSS projects, most research focuses on selecting a manageable subset of projects from \textcolor{black}{the largest ``forge'', GitHub}. Once projects are selected according to some criteria, \textcolor{black}{ data
%from
of these projects} are extracted, cleaned, and analyzed to identify a relationship between some predictors and outcomes investigated in specific research~\cite{tutko2020effective, 10.1145/2597073.2597074, 9463094}. In fact, in more than half of the cases of such analysis obtained from leading conferences, the projects were selected based on some measure of popularity. \textcolor{black}{Another study~\cite{10.1145/2901739.2901776}, investigated methods used in 93 publications that analyzed the mining of GitHub repositories with respect to reported methods, datasets, and limitations and found numerous issues with the dataset collection process and size and poor sampling techniques.} The authors of~\cite{7816479} filtered the top 2500 public repositories on GitHub with high stars. The goal of this restriction was to concentrate on the peculiarities of the widely used GitHub systems and the difficulties that come with human examination. The ``star'' ratings are a static representation of the community's preferences at a certain moment in time (namely, the past) rather than current choices. It also does not show which packages are used the most or which are growing or losing popularity. In \cite{Linstead2009, 5693210}, the authors proposed ranking the code and filtering the repositories based on the entity of the code and its usage. This approach is based on \textcolor{black}{$\sim$12,000 Java projects} filtered by the presence of source code, resulting in the database consisting of 4600 repositories. Such filtering can surely exclude projects that are widely used and safeguards against phishing, intentional code spoofing, among others. Along the same line \cite{10.1145/1882291.1882316}, the authors performed a filtering based on the static snapshot of code repositories based on one-time usage, resulting in more than 200K filtered code bases. In~\cite{6976089}, the authors illustrated the relationship between popularity and security level based on the 
%the top 20 Java libraries widely used. 
\textcolor{black}{top 20 widely used Java libraries.} The filtering was based on usage rather than on more elaborate and finer-grained experiments. \textcolor{black}{Research has shown that in open source software projects, factors like the number of branches, open issues, and contributors significantly influence popularity, with each factor having a large effect size~\cite{8754436}. Another study indicates that the number of followers and aspects such as programming language and commits by others at the time of newcomer joining are crucial for long-term contributor retention, highlighting these as top features in various time intervals~\cite{8721092}}. 
%\textcolor{black}{The practice of substituting professional developers with convenience samples in security developer studies, as highlighted in~\cite{205176}, raises significant concerns regarding the representativeness and validity of the findings in this field.} 
 Several research studies have highlighted the use of collaborative filtering as a way to identify key repositories to examine a multitude of scientific problems \cite{SUN2020106163, 10.1007/978-3-030-03596-9_28, 4052914}. Such studies clearly raise the problem of trust, reproducibility, and validation resulting from incomplete data samples through heuristic filtering of open source repositories. 

%Some work provides no reasons for such filtering, presumably implying that it is the most sensible and widely applied practice needed to a) weed out insignificant software projects and non-software related repositories, and b) get manageable but representative (of real software projects) data. 
\textcolor{black}{Authors in ~\cite{7887704} identify the use of small data sets, poor sampling techniques and hard to replicate methodologies, while authors of~\cite{10.1007/s10664-017-9512-6} recognize that contemporary filtering methods may be
insufficient means of identifying ``real'' software projects and
trie to use machine learning models that rely on various repository
statistics to identify ``real'' software projects.}  \textcolor{black}{In particular, while our proposed work focuses on the methodological robustness of sampling strategies over time,~\cite{10.1007/s10664-017-9512-6} offers a practical tool for filtering engineered projects in large software repositories.  Thus,~\cite{10.1007/s10664-017-9512-6} and our proposed work address limitations in current research practices but from different angles: one through methodological improvement and the other through a technical solution.
}

\textcolor{black}{In summary, some}
measure of popularity or project activity is often used to filter out
projects that are below certain thresholds for that measure or for prediction probability (in the case of ML models). As a result, many software projects are excluded from the subsequent
analysis based on filtering criteria, resulting in the so-called ``missing
data''. 

\textcolor{black}{Missing data, if not handled properly, distorts statistical results and degrades the prediction of machine learning methods. Filtering removes the response and all independent variables, while traditional missing data techniques are typically used when at least one of the predictors is present and uses imputation and/or weighting techniques to fill in missing values~\cite{little2019statistical,Mockus2008}. The three types of missing data~\cite{little2019statistical}  are MCAR or ``missing completely at random'', MAR or ``Missing at random,'' and MNAR or ``missing not at random.'' Filtering would result in MCAR if the distribution of the response is independent of the filtering criteria, a quite unlikely scenario in most cases. The advantage of MCAR is that the analysis on the filtered data would be valid. Fortunately, the analysis would be valid even in MAR cases, which require that the missingness be conditionally independent given the observed variables, i.e., the fact that a value is missing depends only on the observed values. Since filtering excludes all values, subsequent analysis is equivalent to the listwise-deletion technique (i.e., when data rows with any missing values are excluded). It yields correct results only if the filtering criteria are completely unrelated to the response: an extremely unlikely scenario.  } 

% \sout{Without getting into details, such
% filtering represents a ``missing not at random'' (MNAR) type of problem that cannot be ignored, as it can dramatically change the results of the analysis. Specifically, in this case, the filtered data may not be ``missing at random'' (MAR): a
%   critical requirement to obtain sound analysis results when some
%   data have been excluded (see, e.g.,~\cite{Mockus2008}).
%   Filtering, in essence, excludes certain data from a model. It
%   turns out that if the sampling is random, such exclusion should
%   not bias the results systematically. Such case is referred to as
%   missing completely at random (MCAR), i.e., the fact that the data
%   observation is absent is probablisticaly independent from the
%   value of that data point.  The MAR assumption is less restrictive
%   as it allows the probability that a datum is missing to depend on the
%   datum itself indirectly through quantities that are observed.  
%   }
  \textcolor{black}{
  For example, if we predict the number of stars, and the likelihood that the number of stars is missing 
  depends only on the number of authors, we need to include the number of authors in a regression model and impute the missing values for the number of stars to get valid results. Filtering, unfortunately, excludes entire data rows where the number of authors is low, hence any subsequent analysis would be invalid.} 
  MCAR assumption would not apply, because the probability
  that a value is missing depends on the number of authors.
  However, if we do not have the number of authors (in case of filtering) or simply do not
  include it in our estimation model, then even the MAR assumption
  is not satisfied.  Such case is referred to as data not missing at
  random (MNAR).  The MNAR data can be made to satisfy the MAR
  assumption if variables that characterize situations when a value
  is missing are added. Therefore, it is important to add variables
  that might predict the missing value mechanism to the dataset. With filtered data that is not possible.

In this study, we illustrate the often overlooked issue of missing data
arising when software projects are filtered out of research
datasets. We demonstrate the significant impact this can have on
research outcomes, particularly through a case study that models the
popularity of software projects. 

Our recommended approach to solving
this problem involves either using the full data when possible (research infrastructure such as World of Code~\cite{ma2020world} and archival collections such as 
 \textcolor{black}{ Software Heritage~\cite{software-heritage} make complete datasets much easier to access) or the use of a stratified sampling technique~\cite{neyman1992two} from the above-mentioned resources to }
effectively reduces the risk of bias, ensuring that the empirical
findings remain representative despite excluded (missing) data. \textcolor{black}{Our proposed use of stratified sampling might help addresses the concerns highlighted in~\cite{Baltes2022}, which emphasizes the rarity of representative sampling in software engineering research, proposing a solution that contributes to resolving the field's generalizability crisis.
The key idea of the stratified sampling is that different populations (strata) of projects may have different properties based on certain criteria, for example, the mechanism that gets a single-author project many stars may be different from one for projects with thousands of contributors. Instead of getting data on all projects, a random sample is selected for each strata and then the results are extrapolated to the entire population. A simple random sampling would also work, but it would require a much larger sample size for the same variance of the desired estimates, e.g., a very large random sample is needed to include at least some of the very large (and extremely rare) projects.}
%What sets our research apart is not just the breadth of our analysis, which is more extensive than typical studies, but also our %pioneering
%application of stratified sampling in this context, enhancing the credibility and applicability of our results in the field of software engineering.

  \textcolor{black}{
 First, we explained how filtering is related to missing data and how analysis on filtered data is equivalent to listwise deletion technique. Second, we provide an actual example based on a very large real dataset that demonstrates that filtering may completely change the conclusions. Third, we illustrate how stratified sampling could be applied to manage the scale of the analysis yet yield representative results.}

%The contributions of this paper include raising the problem of missing data resulting from filtering out software projects; illustrating how the problem affects the results of analysis based on a simple illustrative example predicting the popularity of a software project;  proposing ways to obtain valid estimates by conducting stratified sampling of software projects that ensure that the missing data from excluded projects will not bias the findings; and our contribution lies in the extensive scale of our analysis and the application of a stratified sampling methodology that is not commonly employed in the field.  

We draw data from a collection of open source version control repositories that attempts to approximate the complete universe of open source \textcolor{black}{ software, World of Code (WoC)~\cite{ma2020world},} in order to be able to draw a truly random sample and compare it with filtering approaches commonly used in empirical software engineering research.  

The remainder of the paper introduces an illustrative example in Section~\ref{s:ill}, discusses limitations in Section~\ref{s:lim}, discusses stratified sampling in Section~\ref{s:str}, and concludes in Section~\ref{s:con}.

\textcolor{black}{
All data and code used in this study are publicly available in the replication package links: 
\url{https://zenodo.org/records/10516681} and \url{https://github.com/woc-hack/Assignments/}}

\vspace{2mm}
\section{Illustrative Example}\label{s:ill}
\label{sec:modelstar}
% \vspace{1mm}
This section 
% describes our proposed approach to 
illustrates the problem of changes in the rating resulting from missing and filtering the repositories based on certain attributes. To model the rating of software repositories and measure what causes the ratings to change, we rely on key inherent characteristics of the repositories and model these quantities to understand what causes some projects to have more stars than others. 

\subsection{Problem statement}
%and method}
% \vspace{1mm}
We would like to explain the popularity of a GitHub repository by various other attributes. Our primary hypothesis is that repositories with more activity, more authors, and those that have been around for a longer time will be more popular. To operationalize these concepts, we use the number of stars as a measure of popularity, the number of commits as a measure of activity, the time since the repository was created as duration, and the number of individuals who committed code to the repository as the number of authors. 

For simplicity, we used linear regression to model the number of stars, and the following sections describe precise ways in which measures were obtained and the sample was filtered.

\subsubsection*{Data Extraction}
Since its inception, Free/Libre and open-source software (FLOSS) has dramatically impacted the software community and ecosystem. FLOSS is free software and open source software, so anyone can freely use, copy, study, and change the software in any way. World of code (WoC) has enabled research on the global properties of \textcolor{black}{FLOSS} %Free and Open Source Software (FOSS) 
\cite{ma2020world}. The WoC collection of software repositories contains authors, projects, commits, blobs, dependencies, and a history of the FLOSS ecosystems. These data are updated regularly and contain billions of git objects. We will use WoC's extensive and frequently updated version control commit data collection of the entire FLOSS ecosystems for this study. WoC APIs are used to fetch and process data related to
\textcolor{black}{FLOSS} 
repositories. We extract several attributes of repositories' to develop the ranking\textcolor{black}{/popularity rating} mechanism. To baseline the analysis and develop a control dataset, we have extracted more than 100K repositories containing the following attributes about each repository. 
\begin{itemize}
    \item project id (\textit{p})
    \item the starting date (\textit{fr})
    \item project number of stars (\textit{ns})
    \item the project duration (\textit{dur})
    \item number of authors (\textit{na})
    \item number of commits (\textit{nc})
    \item number of Core Developers (\textit{nCore})
    \item number of commits by Top Developers (\textit{nc1})
\end{itemize}

Please note that the quantities obtained are highly skewed; hence we do a log transform to make it more suitable for linear regression. Next, we begin by performing a correlation analysis on the extracted data to estimate the relationship among the attributes of the repositories. Understanding the correlations among the predictors informs the interpretation of the results of the model. To simplify the interpretation, we typically want to leave only one of the highly correlated predictors in the model~\cite{M14}. 
% \usepackage{graphicx}
\begin{table}[]
\caption{Correlation analysis based on collected data and analysis results using Spearman correlation method.}
\label{tab:corr}
\resizebox{\columnwidth}{!}{%
\begin{tabular}{|l|r|r|r|r|r|r|}
\hline
 & \textbf{ns} & \textbf{dur} & \textbf{na} & \textbf{nc} & \textbf{nCore} & \textbf{nc1} \\ \hline
\textbf{ns} & 1.00 & 0.26 & 0.16 & 0.17 & 0.08 & 0.15 \\ \hline
\textbf{dur} & 0.26 & 1.00 & 0.14 & 0.25 & 0.06 & 0.16 \\ \hline
\textbf{na} & 0.16 & 0.14 & 1.00 & 0.28 & \textbf{0.70} & 0.06 \\ \hline
\textbf{nc} & 0.17 & 0.25 & 0.28 & 1.00 & 0.16 & \textbf{0.89} \\ \hline
\textbf{nCore} & 0.08 & 0.06 & \textbf{0.70} & 0.16 & 1.00 & -0.10 \\ \hline
\textbf{nc1} & 0.15 & 0.16 & 0.06 & \textbf{0.89} & -0.10 & 1.00 \\ \hline
\end{tabular}%
}
\end{table}
\vspace{0.5mm}
\subsubsection*{Correlation Analysis}
We begin with a correlation analysis of the independent variables extracted for the repositories to measure the strength of the linear relationship. This allows us to determine how much one variable changes due to the change in the other. A high correlation indicates a strong association between the two variables, whereas a low correlation indicates a poor relationship between the variables. We have used Spearman's correlation for this study. 
%\todo {check ont his, why did we inlcude 0.28, i think it was ment to be .70 and .89} 
The results of the analysis shown in Table \ref{tab:corr} illustrate a strong correlation 
\textcolor{black}{between the number of authors and number of core developers,}
and the second
%between the number of authors and commits, and the second
between the number of commits and the number of commits by the top developers,
% \textcolor{black}{ the number of commits and the project duration and the project number of stars and the project duration.} 
This conclusion would help us to understand the interdependence of random variables in a large data set of the repository. Next, we define our approach to calculate the popularity as a function of the repository's set of independent variables. 

In this section, we demonstrate how filtering can alter the calculation of the \textcolor{black}{popularity rating} of open source repositories. The meaning and calculation of the popularity vary based on the underlying formulation and the definition of the popularity itself. For example, some studies define popularity as the download of software among the practitioner community. In comparison, some use developer activities as a function of assigning popularity to open source repositories \cite{5693210}. Others have used HITS-based influence analysis on graphs that represent the star relationship between Github users and repositories \cite{Hu2016}, among others \cite{6976089, 7816479, 1423993}. However, whatever definition entails, we demonstrate how a chosen well-defined popularity rating of software can change when a varying number of repositories are studied.

\subsubsection*{Popularity rating}
\vspace{0.5mm}
The software repository rating is measured as a linear combination of log of duration (\textit{ldur}), number of commits (\textit{lnc}), number of authors (\textit{lna}), and number of core developers (\textit{lncore}).  We begin by defining a plausible rating definition as a combination of several \textcolor{black}{repository attributes}. Then we show whether such a \textcolor{black}{set of attributes} is a reliable mechanism to estimate the ratings. We propose the use of these attributes to estimate the ratings\textcolor{black}{/rankings} of repositories. However, in doing so, we also demonstrate the argument that repository ratings can vary as we modify the definition and the contribution of repository attributes in calculating the ratings. That said, let us assume we define multiple linear regression as follows, 

\vspace{0.5mm}
$ lm(lns \sim ldur+lnc+lna+lCore, data=za)$

\vspace{1mm}
To determine the repository \textcolor{black}{popularity rating}, we employ multiple regression, is a statistical technique for predicting the outcome of a response variable by combining several explanatory variables. The linear relationship between explanatory (independent) and response (dependent) variables is represented using multiple linear regression.

Multiple regression is essentially an extension of ordinary least-squares regression because it uses more than one explanatory variable. We utilize this method to calculate the \textcolor{black}{popularity rating} using a linear combination of the attributes mentioned above. In addition, the relationship between many independent criteria and the calculation of the dependent \textcolor{black}{popularity rating} is investigated.

After each of the independent factors has been determined to predict the dependent attribute, the information on the numerous attributes can be used to provide an accurate prognosis on the level of effect they have on the outcome of the \textcolor{black}{popularity rating}. Next, we begin with an unfiltered random sample of repositories that serves as a control (baseline) for the repository ranking. Then, we show how filtered out or missing repositories from this control sample can alter the rating calculation as the estimates of the independent variables change. This illustrates that changes in filtering result in a change in the outcomes of the analysis.  

\vspace{0.5mm}
\subsubsection*{Control Sample}
We begin with the creation of a control sample of 100k repositories to serve as the baseline estimate for the independent variables contributing to the ranking of the repository. The control data will remain unchanged throughout the experiment. It will be used as a reference and for comparison with other filters that induce missing data and evaluate potential changes in the dependent variables. We test whether the independent variables (predictors) that include log of duration (\textit{ldur}), log of number of commits (\textit{lnc}), log of number of authors (\textit{lna}), and log of number of core developers (\textit{lncore}) go from positive (or reverse) on the dependent variable (project number of stars (\textit{ns})).

We show the results of the multiple regression analysis of control data in Table \ref{tab:allins}. The commits made by the developers (\textit{lnc}) has an estimated value of 0.0232. Next, we examine the effect of filtering by the number of authors (\textit{lna}). 

\begin{table}[]
\caption{MLR with random control sample of 100K repositories}
\label{tab:allins}
\begin{tabular}{|l|l|l|l|l|}
\hline
\textbf{Attribute} & \textbf{Estimate}           & \textbf{Std. error} & \textbf{t-value} & \textbf{Pr(\textgreater{}|t|)} \\ \hline
ldur & 0.155 & 0.0046 & 33.600 & \textless 2e-16 \\ \hline
lnc  & 0.023 & 0.0070 & 3.100  & 0.0019         \\ \hline
lna  & 0.795 & 0.0240 & 32.800 & \textless 2e-16 \\ \hline
lnCore             & -0.564 & 0.0350            & -16.126          & \textless 2e-16                \\ \hline
\end{tabular}
\end{table}

%\subsubsection*{Projects with 300+ authors}
%In order to demonstrate the rating, we select a sample of projects containing at least 300 authors of more. Next, we calculate the MLR to estimate the star ratings through the independent variables ($ lm(lns \sim ldur+lnc+lna+lCore)$). The results of this shown in Table-\ref{tab:allins2300}. These results shows that commit is negatively influenced by the number of commits, when the number of authors reduced. The commit value in the control dataset  is positive while the calculated value in this case for 300+ authors became negative. This result demonstrate our proposed motivation that filtering changes the empirical studies.  

\vspace{0.5mm}
\subsubsection*{Projects with 10+ authors.}
To illustrate by how much the filtering may alter the results, we select a sample by filtering the list of repositories containing at least 10+ authors or more. Since the number of stars is our response variable, any filtering by the number of stars would immediately bias the results. Instead, we filter by one of the predictors: the number of authors. Since the number of authors is included as a predictor in the model, it should not affect the modeled relationship between the number of authors and the number of stars, but filtering is equivalent to listwise-deletion and for the analysis to be valid an MCAR assumption needs to hold (and it does not).
%If such a relationship exists, the presence of the predictor would make the missing data conditionally independent as long as we do not conduct filtering. On the other hand, if we omit the number of authors as a predictor, it would create the situation of missing not at random. Excluding observations with fewer than ten authors may affect other relationships between the number of stars and the properties of the project that we do not measure in the model. This, in turn, may lead to changes in the modeled relationships.

Next, we estimate the \textcolor{black}{popularity ratings} through the independent variables ($ lm(lns \sim ldur+lnc+lna+lCore)$). The results of this are shown in Table \ref{tab:allins210}. These results show that the number of commits has a negative influence when we filter the repositories by the number of authors. Note that the estimated value of log of the number of commits (\textit{lnc}) in the control dataset results is positive, while the estimated value of \textit{lnc} in this case for filtered repositories with 10+ authors became negative. 
%This result demonstrates our proposed motivation that filtering could dramatically change the outcome of empirical studies even in cases when the filtering criteria (missingness) is included as a predictor in the model to avoid an even more serious missing-not-at-random problem.
\textcolor{black}{This dichotomy in results further underscores the crucial point that even when the criteria for data exclusion, such as instances of missing data, are integrated as predictive variables in the analysis, the process of filtering can substantially alter the outcomes of empirical studies.}
%This integration of missingness as a predictor is an attempt to address potential biases arising from data not missing at random. However, our results suggest that this approach, while mitigating some issues, may not completely resolve the distortions in study results caused by such filtering methods.}
\vspace{1.5mm}
\textcolor{black}{
\subsection{Proposed solution} %statement and method}
% \vspace{1mm}
We propose a practical solution for the classification and selection of open source software using stratified sampling. This approach involves dividing software projects into distinct groups before sampling, ensuring that various project types are represented. It addresses some issues like biases from unmeasured factors and missing data, common in software analyses. By weighting each group according to its prevalence in the real world, stratified sampling requires a much smaller sample size than a simple random sampling yet allows inference on the overall population. %This strategy improves the precision and applicability of our findings, making it a reliable and scientifically robust approach to data filtering in software engineering research. 
Next section discusses this in more details. }
%\textcolor{red}{Todo: add more details about what approach we would take for the illustrated example}

\begin{table}[]
\caption{MLR results for more than ten authors}
\label{tab:allins210}
\begin{tabular}{|l|l|l|l|l|}
\hline
\textbf{Attribute} & \textbf{Estimate} & \textbf{Std. error} & \textbf{t-value} & \textbf{Pr(\textgreater{}|t|)} \\ \hline
ldur   & 0.57  & 0.08 & 6.98  & 2.01e-11 ***        \\ \hline
lnc    & -0.52 & 0.11 & -4.55 & 7.93e-06 ***        \\ \hline
lna    & 3.55  & 0.38 & 9.4  & \textless 2e-16 *** \\ \hline
lnCore & -2.74 & 0.32 & -8.66 & 3.32e-16 ***        \\ \hline
\end{tabular}
\end{table}

\vspace{2mm}
\section{Stratified sampling}\label{s:str}
% \vspace{1mm}
The field of polling has developed sophisticated sampling techniques that minimize the effort needed to obtain a reasonably accurate estimate of the outcome of, for example, an election. Since systems such as WoC make it easy to collect data for a range of properties for nearly all software projects, sampling is not technically needed. But research may still require an effort-intensive calculation of additional predictors or a manual qualitative analysis. As such, the need for sampling is almost always present. A simple random sample of even 1000 GitHub repositories would result in tiny, short-duration, inactive projects. Putting extra effort into obtaining additional measures for such a sample would not be advisable. However, depending on the hypothesis and likely relationships in the data, we may develop sampling procedures that minimize missing-data bias. In all cases, we should not filter by the value of the response variable, whether it be popularity, activity, or team size. Such filtering would almost \textcolor{black}{certainly introduce a missing} not at random situation if we do not measure and include in the model *all* variables that affect the outcome and either include all rows (no filtering) or do a simple random sampling. However, it should be possible to filter by predictors that are included in the model assuming that there are no unmeasured predictors. Unfortunately, commonly used models in software engineering explain a relatively small fraction of the variability in the response, suggesting that many important predictors are not measured. The relationships may also vary according to the size of the predictors, further exacerbating biases introduced by filtering.  As an approach to avoid the problem of simple random sampling, it is advisable to carefully design the sampling so that projects with different values of the characteristics that may affect the outcome are included. If we want to obtain results for the entire population, we should weight each strata according to its prevalence in the original population. The effects can also be investigated and compared among the different strata and help to obtain more easily and meaningfully generalizable results.

\textcolor{black}{For example, in our case we may want to sample an equal number of projects $n$ from the sub-populations of projects with one to $k$ authors. Suppose that the population size at each $k$ is $N_k$. Then we can fit a weighted multiple regression where the weights for the observations representing $k$-th strata would be $\frac{N_k}{n}$.
Generally speaking, regression analysis and other inference methods are somewhat different for stratified sampling, see, e.g., ~\cite{ca4ab59b-26cf-3155-a9ac-c52d9688b474} and the details are beyond the scope of this illustration.
}

\textcolor{black}{
Another important question is how to conduct a random or stratified sampling in the first place. First, existing research mostly ignores projects not on GitHub, including many important and large open source projects. Second, studies are restricted on sampling using GitHub APIs or mostly obsolete and partial public datasets and have to devise filtering criteria based on whats available in such datasets. Fortunately, several efforts such as Software Heritage~\cite{software-heritage} and World of Code~\cite{ma2020world} 
spent significant amount of effort to collect nearly complete collection of open source projects from all public forges. In fact, World of Code research infrastructure provides interested researchers with access and training and, among other features, includes extensive support for random or stratified sampling such as numerous summaries of software projects, authors, and even programming APIs.
}
\vspace{-2mm}
\textcolor{black}{\section{Discussion}}\label{s:lim}
% \vspace{1mm}
This work focuses 
%entirely
on demonstrating the impact of filtering on the results obtained in a simple case of linear or logistic regression. The specific example was chosen to be as simple as possible to avoid many other aspects of statistical analysis that are important for obtaining valid conclusions.   We tried to illustrate only the most simple aspects of sampling and did not attempt to cover more sophisticated and largely unsolved problems, for example, sampling from large graphs or linking multiple data sources~\cite{M14}. \textcolor{black}{By selecting features that do not overlap with the filtering criteria, our aim is to mitigate the risk of distributional bias. In the future, we will explore the identification of a set of characteristics that are demonstrably independent of our filtering criterion for further analysis.} The regression in an example is not an attempt to show any causal connection. All variables may conceivably affect each other. Therefore, it is not meaningful to argue whether the number of authors increases the \textcolor{black}{popularity rating} or whether the reverse may be true. Instead, the model simply shows that the variables are related in the samples considered. \textcolor{black}{The performance of regression models can generally be assessed by looking at the significance of the model coefficients, the general value of the model $R^{2}$, and other diagnostics such as residuals and fit graphs.} It is also important to note that some of the repositories may not even represent software projects but may be used for other purposes, so it may be important to exclude such irrelevant cases. However, the retention of legitimate software projects should be done in a way that avoids the biases described here. This may be done as described in Section~\ref{s:str}.

\vspace{2mm}
\textcolor{black}{
\section{Lessons Learned}
% \vspace{1mm}
In the course of our demonstration,
we've learned valuable lessons that shed light on the intricacies of data filtering and its implications for research in open-source software ranking and selection:
%we have gleaned several important lessons that illuminate the intricacies of data filtering and its implications for research in open-source software ranking and selection:
\\
\vspace{0.5mm}
\textbf{\textit{Impact of Data Filtering:}}
This study highlights the substantial influence of data filtering on research outcomes. The act of filtering data based on specific criteria can have a profound impact on the relationships between project characteristics and the target variable. This often leads to results that diverge significantly from those obtained when using unfiltered data.
\\
%\vspace{0.5mm}
\textbf{\textit{The Challenge of Missing Data:}} Another challenge emerges in the form of "missing data." Filtering projects based on certain criteria can result in incomplete datasets, introducing potential biases into the analysis.
Filtering corresponds to analysis with missing data when all rows with any missing (filtered) value are excluded. We argue that the analysis of filtered data is valid only when the filtering criteria are independent of the response, a condition that is typically not satisfied.
\\
%\vspace{0.5mm}
\textbf{\textit{Biases in Data Filtering:}} It is crucial to emphasize that data filtering should not be contingent on the value of the response variable (e.g., popularity). Instead, it should be guided by predictors included in the analysis model, aimed at circumventing the introduction of biases.
\\
%\vspace{0.5mm}
\textbf{\textit{The Role of Stratified Sampling:}} To address issues arising from data filtering and missing data, we recommend the adoption of stratified sampling techniques. Stratified sampling involves categorizing software projects into distinct groups before sampling, ensuring equitable representation across various project types. This approach enhances the precision and applicability of the research findings.
\\
%\vspace{0.5mm}
\textbf{\textit{Prudent Sample Selection:}} Researchers are advised to exercise caution when selecting samples for empirical studies involving software repositories. Thoughtful sample selection plays a pivotal role in mitigating potential biases introduced by data filtering.
\\
%\vspace{0.5mm}
\textbf{\textit{Ensuring Research Reliability:}} This study underscores the paramount importance of guaranteeing the reliability of research results. Researchers are encouraged to perform diagnostic tests, use cross-validation, and consider longitudinal analysis to validate model assumptions and affirm the robustness and consistency of their findings over time.
}

\vspace{1mm}
\section{Conclusion}\label{s:con}
% \vspace{1mm}
Sampling strategies play a crucial role in data discovery science, as the findings are drawn from the data captured for analysis. In this paper, we illustrate how convenience sampling or filtering by various criteria could affect the investigated relationship, leading to very different results, sometimes opposite to what would be obtained from the unfiltered data. These evaluations help to understand the robustness of the research results and ensure that the conclusions drawn from such studies are reliable and representative of the real world scenarios they intend to model. \textcolor{black}{ In the future,
%we plan to conduct diagnostic tests to verify model assumptions and use cross-validation for a more reliable performance estimate. In addition, we consider 
we plan on
conducting a longitudinal analysis to see if the results hold over time.} This is particularly important if the ratings change over time, as it could affect the relevance and applicability of the research findings.

\section*{Acknowledgements}
The work was partially supported by National Science Foundation awards 1633437, 1901102, 1925615, and 22120429. 

This manuscript has been authored by UT-Battelle, LLC, USA under Contract No. DE-AC05-00OR22725 with the U.S. Department of Energy. The publisher, by accepting the article for publication, acknowledges that the U.S. Government retains a nonexclusive, paid up, irrevocable, worldwide license to publish or reproduce the published form of the manuscript, or allow others to do so, for U.S. Government purposes. The DOE will provide public access to these results in accordance with the DOE Public Access Plan (http://energy.gov/downloads/doe-public-access-plan).

%Sampling and filtering is a popular strategy in software engineering research. For example, researchers pick a sample of projects large enough or with more or less starts, and then perform the analysis for their research question. In this paper, we  argue that any form of filtering or  sampling alters the conclusion, so basically, any published study in which you conduct some sort of sampling and filtering and then come up with a result, has a risk of not being entirely accurate.

%\section{Examples}
%In this section, we discuss existing studies related to the use of filtering of software projects.

%\input{casestudy}

%\section{Discussion}
%What should go in discussion?

%\begin{acks}
%This work was supported by NSF grants 1901102 and 2120429.
%\end{acks}
\balance
\bibliographystyle{ACM-Reference-Format}
\bibliography{bibfile}
\end{document}